\documentclass[%
aps,
amsmath,amssymb,
reprint,%
]{revtex4-1}

\usepackage{graphicx}
\usepackage{dcolumn}
\usepackage{bm}
\usepackage{color}
\usepackage{tabularx}
\usepackage{multirow}
\usepackage[normalem]{ulem}

\usepackage{changepage}

\begin{document}

\title{Translational control of gene expression via interacting feedback loops}
\author{Liang Wang}\affiliation{Division of Mathematics, University of Dundee, DD1 4HN, UK}
\author{M. Carmen Romano}\affiliation{SUPA, Institute for Complex Systems and Mathematical Biology, Department of Physics, Aberdeen AB24 3UE, UK}
\affiliation{Institute of Medical Sciences, University of Aberdeen, Foresterhill, Aberdeen AB24 3FX, UK}
\author{Fordyce A. Davidson}\affiliation{Division of Mathematics, University of Dundee, DD1 4HN\\ Corresponding Author}

\email{f.a.davidson@dundee.ac.uk}
\date{\today}

\begin{abstract}
Translation is a key step in the synthesis of proteins. Accordingly, cells have evolved an intricate  array of control mechanisms to regulate this process.    By  constructing  a multi-component mathematical framework for  translation  we uncover how translation may be controlled via interacting feedback loops.  Our results reveal that this  interplay gives rise to a remarkable  range of protein synthesis dynamics, including oscillations, step-change and bistability.  This  suggests that cells may have recourse to a much richer set of   control mechanisms than was previously understood.

\end{abstract}
\maketitle

\noindent
Control of gene expression refers to the processes by which the production of proteins is regulated by the cell. This is at the heart of the functioning of all living organisms and it allows cells to adapt  to their environment. Control of gene expression can occur at multiple levels. In this Letter we focus on translational control.\\
 Translation is the process by which a protein is made from a messenger RNA (mRNA) molecule.
 An mRNA consists of a sequence of codons, each coding for a certain amino acid. Translation is performed by molecular machines called ribosomes, which bind to the beginning of the mRNA (5' UTR region), scan it for the start codon and hop from one codon to the next, thereby producing the chain of amino acids which  form the protein. When the ribosome reaches  the stop codon, the protein is complete, is released into the cytoplasm and the ribosome binds off the mRNA.\\
Recent years have witnessed an explosion of information about how translational mechanisms regulate protein levels~\cite{translational_control}.  Prominent examples include translational control during cell stress~\cite{Spriggs10}  and switching  in the mechanism responsible for translation initiation during the cell cycle~\cite{Sonenberg01}.

Here we focus on one important case of translational control that has remained 
unexplored within this research framework, namely the interplay between positive and negative regulatory mechanisms. Translational negative feedback is caused by the ability of the produced proteins to bind to the 5'UTR region of their own mRNAs and hinder initiation~\cite{Brunel95,Betney10}. In fact, this is the case for a crucial protein in the cell called PABP (PolyA Binding Protein), which promotes translation initiation and protects the transcripts from degradation enzymes~\cite{Cao01,Hornstein99,Neto95,Bag01}. On the other hand, virtually all mRNAs in the cell are subject to positive feedback via ribosome recycling due to their pseudo-circular structure~\cite{Wells98}, so that terminating ribosomes can be recycled back onto the same mRNA to commence a new round of translation. In this Letter,  we show that the interplay between negative translational feedback and ribosome recycling gives rise to a novel range of dynamical behaviour in protein synthesis, including oscillations, step-change  and bistability. Therefore, mRNAs  for which the ubiquitous  ribosome recycling  positive feedback loop  is augmented by  negative translational feedback are endowed with a mechanism  that allows them to finely tune protein synthesis according to  environmental conditions.\\

\begin{figure}
\center
\includegraphics[width=0.5\textwidth]{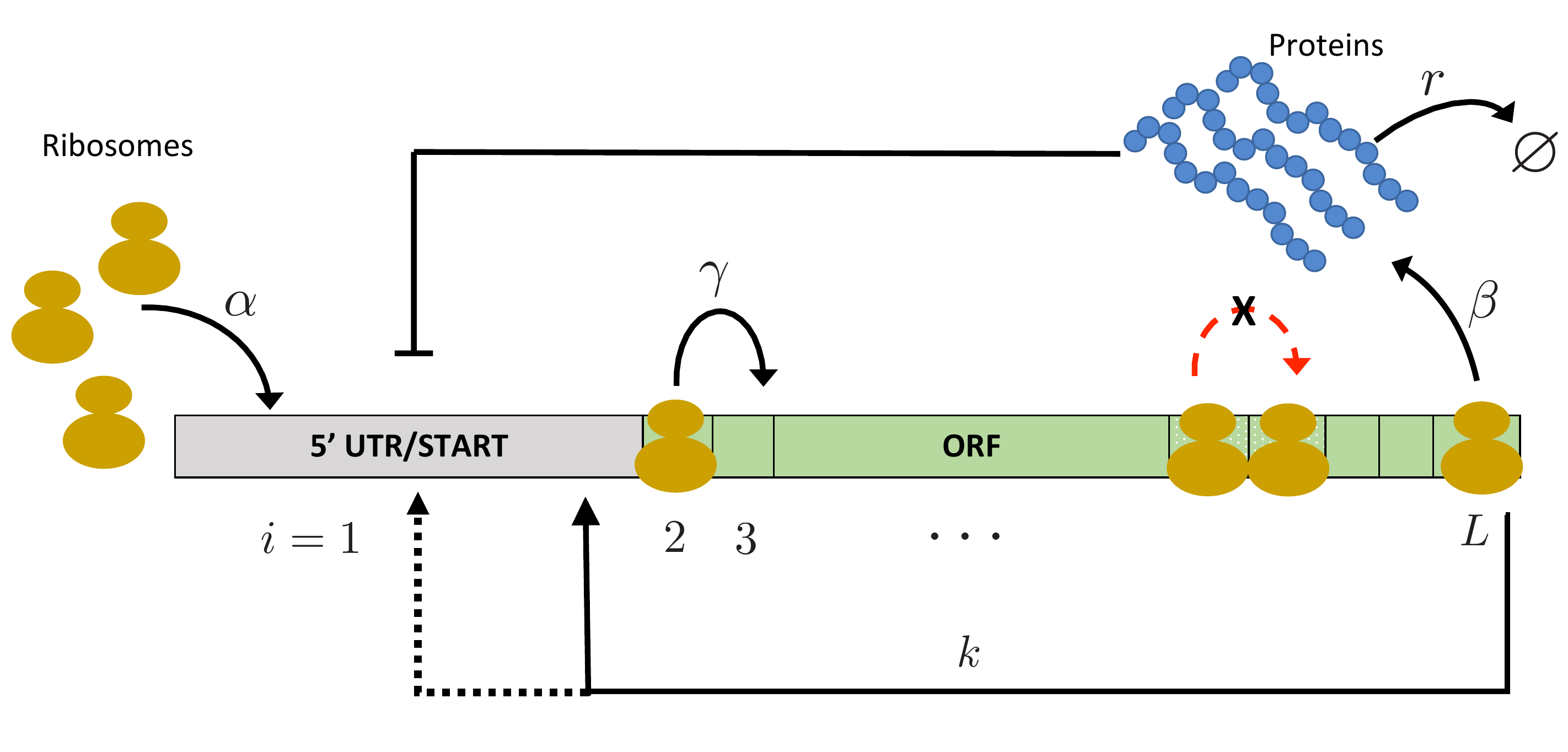}
\caption{Schematic of translation model with ribosome recycling and auto-negative feedback. Competitive recycling  (dashed line),   non-competitive recycling  (solid  line). Each site can be occupied by no more than one particle, so that at any  time $t$ the state  at site $i$ given by  $S_i(t)$   is either $S_i(t) = 0$ or $S_i(t) = 1$, with $ i = 1,...,L$, where  $L$  is  the  lattice length.  Particles bind to the first site of the lattice at rate $ \alpha $, then hop from one site to the next at rate $\gamma$ (usually rescaled to one  and done so here) and finally leave the lattice from the  last site at rate $ \beta $, marking the point where the associated protein synthesis is completed.  See text for further details.}\label{Schematic}
\end{figure}

\noindent 
{\bf Modelling framework.}--Our mathematical framework is a multi-component model that accounts for translation, protein complex formation and  binding  of protein complexes and ribosomes at the 5'UTR (see Fig.~\ref{Schematic}).\\

\noindent
{\em Translation.} We use a stochastic model of one-dimensional transport extensively studied in non-equilibrium statistical physics, called the  Totally Asymmetric Simple Exclusion Process (TASEP)~\cite{Derrida92}. Ribosomes are represented by particles that hop stochastically along the sites of a one dimensional lattice, corresponding to the codons of the mRNA~\cite{MacDonald68}. 
Traffic on the lattice can be classified into three main phases:  the \textit{low density} (LD)  phase ($ \alpha<\beta,\,\alpha<1/2 $), the \textit{high density} (HD) phase ($ \beta<\alpha,\,\beta<1/2 $) and the \textit{maximal current} (MC)  phase ($ \alpha,\,\beta\geqslant1/2 $), limited by the initiation, exit and internal hopping rates,  respectively. These phases have distinct average  density, $ \rho $, (average number of particles per site) and current, $J$, (average number of particles hopping from one site to the next per unit time), which in the limit of an infinitely long lattice are given by: $ \rho_{\scriptscriptstyle LD}=\alpha,\,\rho_{\scriptscriptstyle HD}=1-\beta,\,\rho_{\scriptscriptstyle MC}=1/2 $ and $ J_p=\rho_p\left(1-\rho_p\right)$, $p \in \{LD, HD, MC\}$~\cite{Derrida93,Schutz}.\\

\noindent
{\em Protein degradation.}
Once synthesised, proteins enter the intra-cellular pool,  where they are subjected to degradation. 
Net removal  from the protein pool is therefore reasonably  modelled as a Poisson process with rate $r$. Hence, the  average  protein number in steady state is given by $ N=N^*:= J/r$. \\

\noindent
{\em Translational negative feedback.}
As detailed above,  a protein  can bind  (often in multimeric form) to the 5'UTR of its own mRNA, thereby blocking the loading of ribosomes and thus  repressing  its own translation. 
Since protein 
binding/unbinding to the mRNA  is generally much faster than ribosome loading~\cite{Chen04}, it follows after some analysis that  the probability of the start codon being free for ribosome loading can be described by a Hill-function
$f(N)=1/\left(1+(N/\theta)^n\right)$, where $N$ is the protein copy number, $ \theta $  measures the protein level that induces half maximal  ribosome  binding  rate and $ n $ is the Hill coefficient measuring cooperativity of the protein multimer (see Supp. Mat.). Thus,  the intrinsic  initiation  rate is modified from the standard constant rate, $\alpha$, to $\alpha_F:=\alpha f(N)$. Note that with the process in steady state, $N\equiv N^* = J/r$ and hence, we can write $f= 1/\left(1+\left(4 I\,J \right)^n\right)$, where we have introduced the reciprocal factor $ I:= 1/(4 \theta r) $ that measures {\it feedback intensity} (the factor of 4 is for algebraic convenience).\\

\noindent
{\em Translational positive feedback.}
The two ends of the mRNA can interact leading to a pseudo-circular structure~\cite{Barthelme11}, which together with the recycling complex Rli1p~\cite{Shoemaker11} promote terminating ribosomes to start a new round of translation on the same mRNA~\cite{Amrani08}. Following the approach in~\cite{Marshall14}, a ribosome on site $ i=L $ is assumed to  either detach at rate $\beta$ and enter the reservoir of free ribosomes or move  directly onto site $ i=1 $ at a  {\it recycling rate} $ k $ (if $ S_1(t)=0 $) to re-initiate the translation process. \\

\noindent
{\em Model for interacting  feedback loops.} 
Experimental results suggest that recycled ribosomes are channelled downstream of the normal {\it de novo} initiation site and thus may evade the blocking effect of the protein complex  \cite{Rajkowitsch04}. This is  the case discussed  here and referred to as {\em non-competitive recycling}. However, 
  the relative position  of the  protein complex  binding site  and the recycled ribosome initiation  site   is not clear. Hence,  in \cite{WRD} we consider the alternative that   both recycled ribosomes and {\it de novo} initiation are blocked by the protein complex, ({\em competitive recycling}) and compare the two mechanisms (see Fig.~\ref{Schematic}).
  
   In the {\em non-competitive recycling} case we obtain the following mean-field equations (neglecting correlations between neighbouring sites) for the average occupancies $\rho_i$ of the lattice sites 
\begin{eqnarray}\nonumber
\frac{d\rho_1}{dt}&=&\underbrace{\alpha f(N)(1-\rho_1)}_{\text{{\it de novo} }}+\underbrace{k\rho_L(1-\rho_1)}_{\text{recycled}}-\rho_1(1-\rho_2),\\ \label{non-comp}\\ \nonumber
\frac{d\rho_i}{dt}&=&\rho_{i-1}(1-\rho_i)-\rho_i(1-\rho_{i+1}), \;\;\; i=2, \ldots, L-1,\\ \nonumber
\frac{d\rho_L}{dt}&=&\rho_{L-1}(1-\rho_L)- \beta  \rho_L -  \underbrace{k (1-\rho_1)\rho_L}_{\text{recycled}}.
\end{eqnarray}
By direct comparison with the standard TASEP, {\it effective} entry and exit rates can be defined as follows
\begin{equation} \label{noncomp_rates}
\alpha_{eff}:=\alpha f(N) +k\rho_{L},\qquad \beta_{eff}:=\beta+k(1-\rho_{1}).
\end{equation}
{\bf Steady-state analysis of protein production and ribosome density.}--Applying  the mean-field TASEP approach to our model by  imposing the conditions for each of the characteristic phases (LD, HD or MC) to the effective rates $\alpha_{eff}$ and $\beta_{eff}$ leads to the derivation of analytical expressions for the protein production rate $J$ and ribosome density  $\rho$ for each phase, as well as the phase boundaries  (see Supp. Mat.). 
These analytical expressions agree well with Monte-Carlo simulations (see Fig.~S1).  This first  analysis  suggests  that  the long term dynamics of our model are unaltered from those of the standard translation model.  However,  a deeper analysis reveals that the complex interplay of the positive and negative feedback  generates entirely novel dynamical responses. \\

\begin{figure}
\centering
\includegraphics[width=0.5\textwidth]{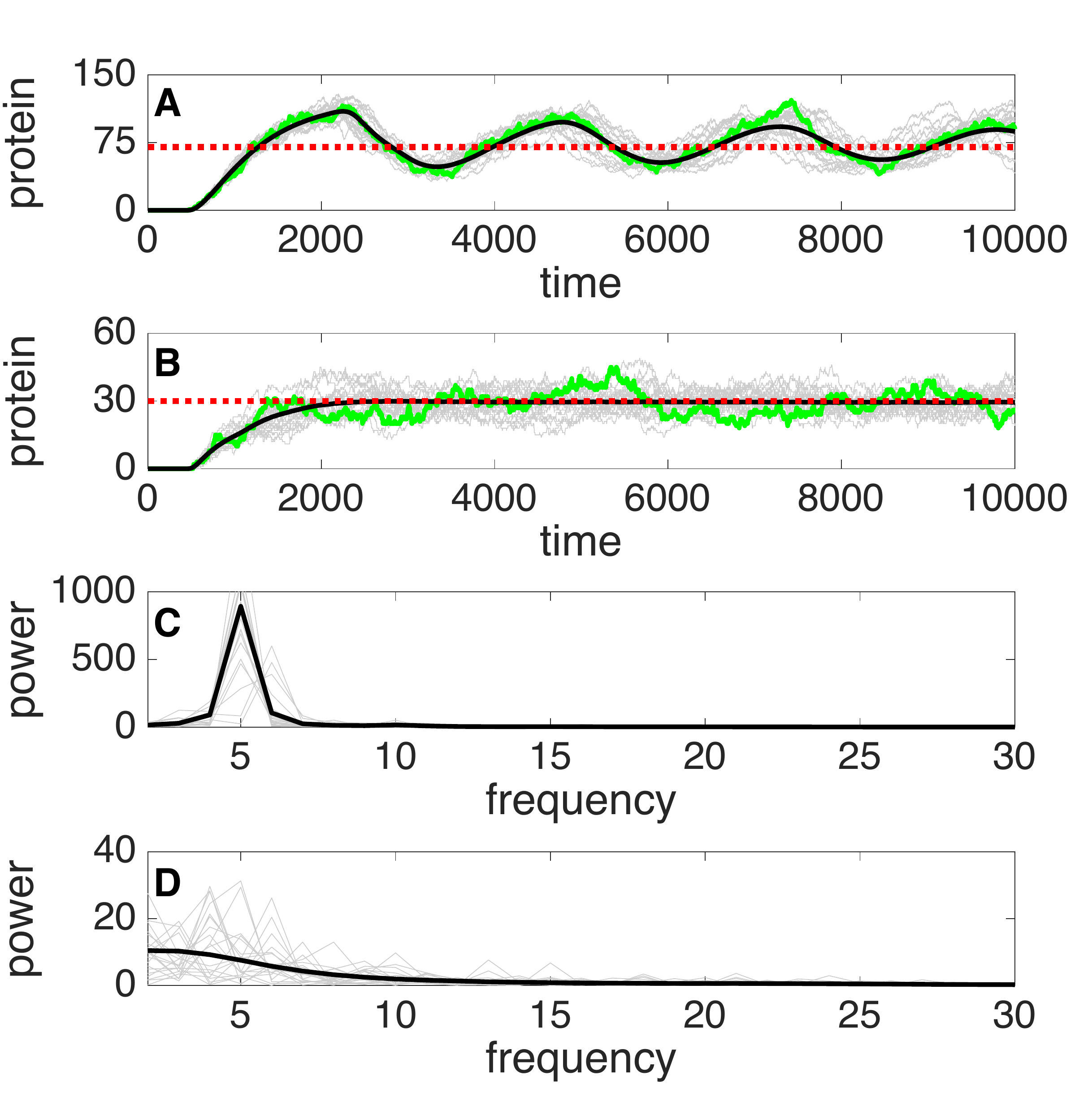}
\caption{Simulation and power spectrum of protein level in low density phase. (A,B) Protein number as a function of time. Averages over 5,000 realizations of stochastic simulations (black lines),  20 example realizations (grey lines), single example trajectory (green line). Red dashed lines are mean field $ N^\ast $ as computed using the steady state theory. Typical simulation (A) and the corresponding power spectrum (C) from the region where oscillation are predicted to exist  ($\alpha=0.8$). Typical simulation (B) and the corresponding power spectrum (D) from the region where oscillation are not predicted to  occur ($ \alpha=0.05$). In all cases $\beta=0.5,\; k=0.2,\; I=10, \;r=0.002,\;n=5,\;L=500$.}\label{Fig-spectrum}
\end{figure}

\noindent
{\bf Negative feedback and ribosome recycling induce oscillations in cellular protein level.}--
Monte-Carlo simulations reveal periodic oscillations in the number of proteins $N(t)$ within the initiation limited regime (LD phase). The stochastic nature of the individual simulations leads to slight fluctuations in the period of the oscillations making it  difficult to systematically differentiate periodic oscillations from random fluctuations by visual inspection. However, a power spectrum analysis provides a  clear demarcation: a  tight, single-peaked spectrum is associated with the apparent  periodic oscillations (Figs.~\ref{Fig-spectrum}A,C) whereas a broad band response is obtained in the case of stochastic fluctuations (Figs. ~\ref{Fig-spectrum}B,D).

The appearance of oscillations  is not wholly unexpected: the time needed for a  ribosome  to transit the mRNA naturally generates a delay between initiation and completion of protein synthesis.   This generates a delay in the action of the negative feedback, a mechanism commonly known to generate  oscillatory behaviour ~\cite{negat_feedback}. 
To get a better understanding, we formulated a simplified model for the  protein copy  number $N(t)$ in the LD regime:
\begin{eqnarray}\nonumber
\dfrac{dN(t)}{dt}&=&J(t) - rN(t)\\
&=&\alpha_{\scriptscriptstyle eff}(t-T) \big(1-\alpha_{\scriptscriptstyle eff}(t-T)\big)-rN(t),\label{DDE}
\end{eqnarray}
where $T$ denotes the translational delay time~\footnote{Note that in agreement with numerical simulations, we only expect oscillations within the LD phase, since in both HD  and MC, the current $J$ is independent of the delay between loading and exit (see Supp. Mat.).}.   Using Eq.~\ref{noncomp_rates} and $\rho_L=J/\beta_{eff}$, it follows that 
\begin{equation}
\alpha_{\scriptscriptstyle eff}(t)=\dfrac{\alpha(\beta+k)}{\alpha k+\beta\Big(1+\big(4 Ir N(t)\big)^n\Big)}\label{alphat}.
\end{equation}
Substituting \eqref{alphat} into \eqref{DDE} results  in a delay differential equation for $N$. The translational delay can be estimated as $T = L/(1-\rho)$, where $\rho=\alpha_{eff}(N^*)$. This simplified model reproduces the amplitude and period of the stochastic simulations (cf. Fig.~\ref{Fig-spectrum} A and Fig. S3D). Importantly, \eqref{DDE} is amenable to a stability analysis that allows us to identify conditions for the onset of oscillations. Indeed,  it can be shown that  on increasing $\alpha$ the steady state of \eqref{DDE} can be driven unstable via a  Hopf bifurcation (see Supp. Mat.). After some algebra, it follows that the Hopf locus
is an implicit expression of the form 
\[
B \cos(\sqrt{B^2 -r^2}\, T)+r = 0,
\]
where $B$ is a function of the system parameters. This locus forms a curve in the $\alpha-\beta$-plane (see Fig.~S2). After some algebra, it can be shown that  necessary conditions for the existence of the Hopf locus are  $n>1$ and $I >  F(\alpha, \beta, k,n),$ were $F$ is a positive valued function of the parameters \cite{WRD}.  The condition $n>1$ indicates that cooperativity in protein binding is necessary for the onset of oscillations, in accordance with~\cite{Elowitz00}. The second condition indicates that the onset of oscillations  occurs when the feedback intensity is sufficiently strong.

As we increase the feedback intensity, the Hopf locus shifts  left in the $\alpha-\beta$-plane, indicating that onset of oscillations occurs  at lower values of the intrinsic loading rate $\alpha$ (see Fig.~S2A), as one would intuitively expect. Interestingly, the Hopf locus also shifts  left on increase the recycling rate (see Fig.~S2B). Hence, counterintuitively, ribosome recycling - a positive feedback mechanism  -   {\em enhances} the onset of oscillations~\cite{Ananthasubramaniam14}.\\

\begin{figure}
\centering

\includegraphics[width=0.5\textwidth]{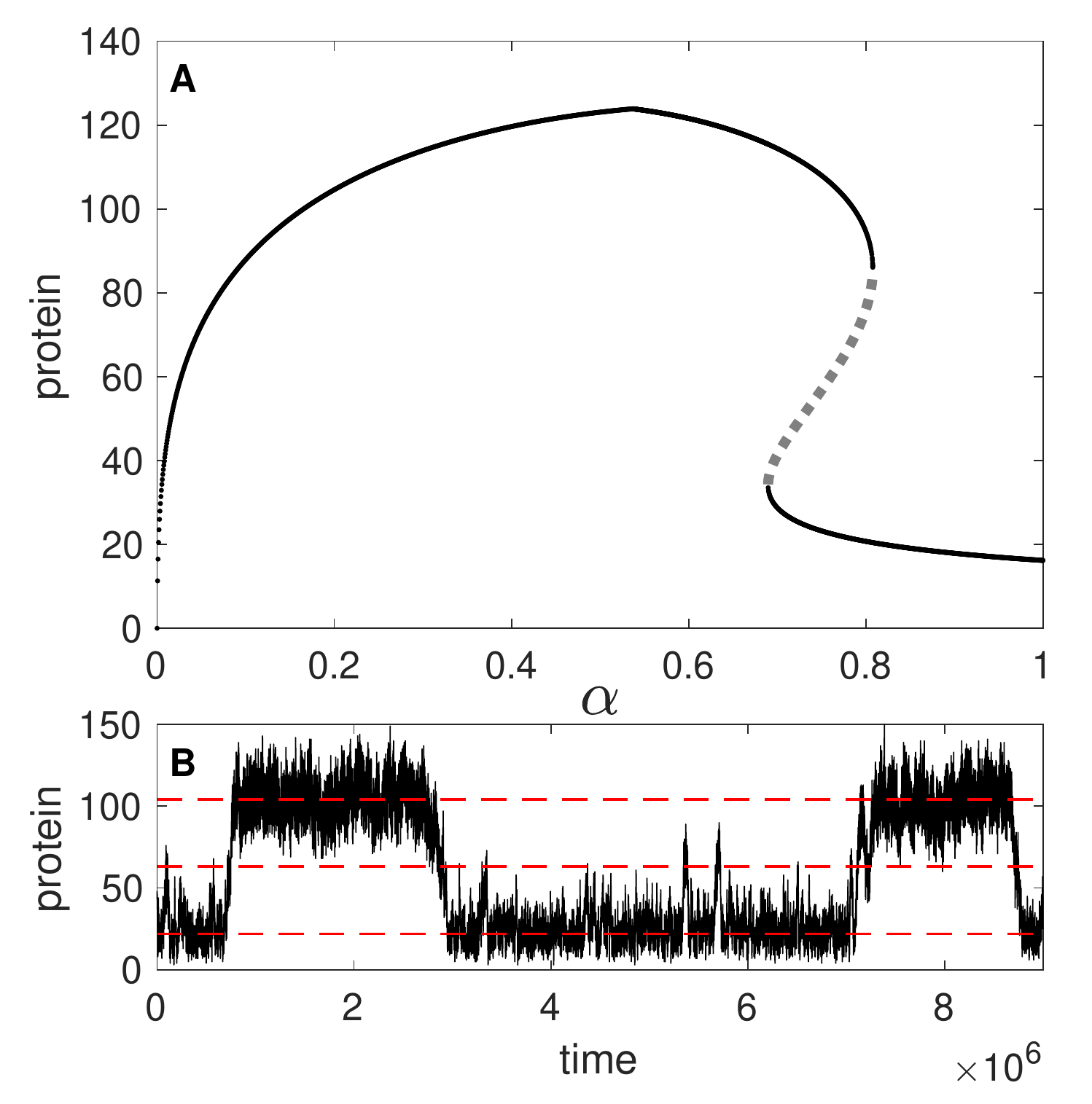}\\

\centering
\caption{ (A) Protein number $N^*$ as a function of  the initiation rate $ \alpha $ as predicted by the steady state theory.  (B) Monte-Carlo simulation of protein number a function of time (solid black line) with the  mean field solutions $N^*$ (red dashed lines) from A  ($\alpha = 0.77$).  In both cases $\beta=0.015$, $k=0.8$, $I=24$,  $r=0.002$, $n=2$,  $L=500$.}\label{Fig4-bistable}
\end{figure}

\noindent
{\bf Interplay between recycling and  negative feedback induces bistability in protein production.} --In the  MC  and LD phases, the current $J$ (and hence $N^*$) is uniquely defined for any given parameter set. On the contrary, in the HD phase  $J= \beta_{eff}\left(1-\beta_{eff} \right)$ and  after some algebra  it can be shown that $\beta_{eff} $ is the solution to the following $ 2n+1 $ degree equation (see Supp. Mat.):
	\begin{equation}
	4^n kI^n\beta \beta_{eff}^n\left(1-\beta_{eff}\right)^{n+1}-(k\beta+\alpha)\beta_{eff}+\beta(\alpha+k)=0.\label{bistable}
	\end{equation}
In the absence of recycling ($k=0$), Eq.~\eqref{bistable} has the unique solution $\beta_{eff} =\beta$. In the absence of negative feedback ($I=0$), $\beta_{eff}$ is uniquely defined by $\beta_{eff}= \beta (\alpha + k)/(k \beta + \alpha)$~\cite{Marshall14}.  However, when both $k,\, I >0$, Eq.~\eqref{bistable} can have three  admissible  solutions, depending on the value of $\alpha$ (see Supp. Mat.). Thus, for suitably chosen parameters, there exists an interval of values of $\alpha$ for which  three steady state values of $N^*=J/r$ co-exist. Figure~\ref{Fig4-bistable}A shows $N^*$  as function of  $\alpha$, so that on  increasing $\alpha$ from zero to one, the model  transits from the LD to the HD phase. $N^*$   is first  monotonically  increasing (LD phase). Then,  at the LD/HD interface, $N^*$ reaches a maximum, and within the HD phase $N^*$ starts decreasing -  the counterintuitive consequence  of ribosome recycling as reported in ~\cite{Marshall14}. As $\alpha$ is further increased, $N^*$  passes  through two fold bifurcations, leading first  into, and  then out of the interval of co-existent states - with the upper and lower branches  separated by an intermediate,  state indicated by the grey dashed line. 

\begin{figure}\label{Fig5-crit-discont}
\centering
\includegraphics[width=0.5\textwidth]{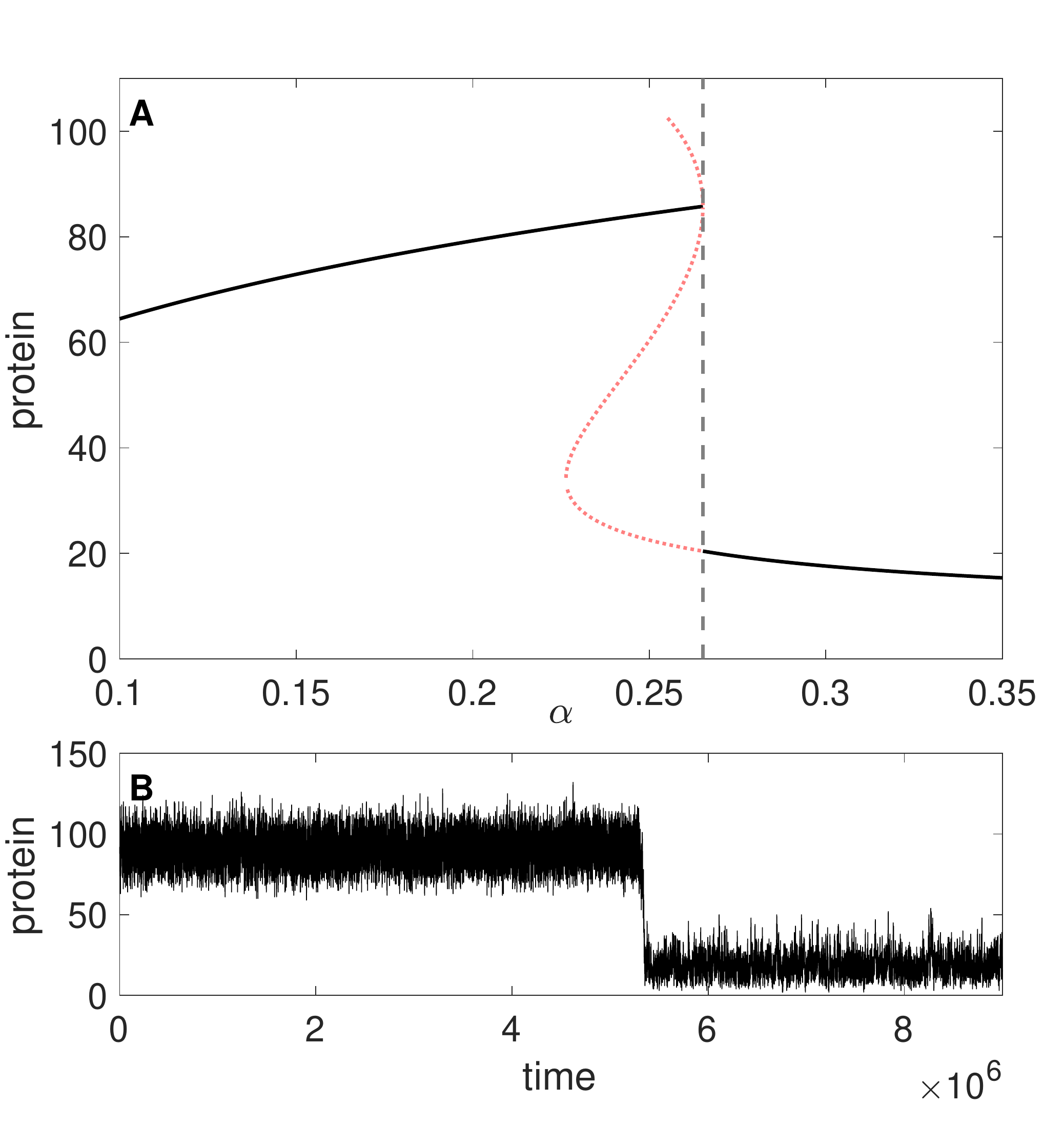}\\
\centering
\caption{Step-change  in steady state protein levels. (A) Signal-response curve for  protein number $N^*$ as a function of  the initiation rate $ \alpha $ (black curves). Inadmissible  solutions for \eqref{bistable} red dotted curve. The LD-HD boundary vertical  grey dashed line.   (B) Time series of the number of proteins:  $\alpha=0.28$ then switched to  $\alpha=0.3$ at time $t=4.5 \times 10^6$. In each case  $\beta=0.015$, $k=0.263$, $I=24$,  $r=0.002$,  $n=2$, $L=500$.
}\label{Fig5-crit-discont}
\end{figure}

With  values of $\alpha$ selected from the co-existence interval, simulations reveal that  the time series  of the number of proteins fluctuates about the high or low state on a time scale orders of magnitude larger than that of the  fluctuations themselves.  Rapid switching  events between the favoured state are   accompanied by a brief hiatus at the intermediate state.   The mean locations  of  these favoured  and intermediate states are well-approximated by the analytic expressions  for the steady states obtained from Eq.~\ref{bistable} (Fig.~\ref{Fig4-bistable}B). Frequency histograms reveal the effect of varying $\alpha$ across the bistable region and together with dwell-time histograms indicate this  to be  a  memory-less stochastic  switching process (see Figs. S5 and S6)

Fixing $k$ (resp. $I$) and increasing $I$ (resp. $k$) increases the interval of values of $\alpha$ for which the fold exists (fold width - see Fig.~S7). Interestingly, the location of the fold is also an increasing function of $I$ and $k$. Indeed, somewhat counter-intuitively, for a fixed value of $\alpha$, increasing the intensity of the negative feedback $I$, can force the system from a low $N^*$ to a high $N^*$ state. To understand this it is important to remember that the bistable region is located within the HD phase. Within the HD regime, any change of parameters leading to a decrease in the ribosome density leads to an increase in the ribosomal current (the lattice is less crowded and therefore particle flow is more efficient). As we increase the intensity of the negative feedback, the effective initiation rate $\alpha_{eff}$ decreases, thereby decreasing the density of ribosomes on the mRNA. Finally, we note that bimodality in protein production rate is a result of the balance between the negative and positive feedback loops and tuning one or the other can drive the system both into and out of a  bimodal response (see Fig.~S7). \\

\noindent
{\bf Feedback interplay can induce step-changes in protein production}--If we now fix the value of $k$ so that bistability is ensured, then  a critical value of $I$  exists at which the right boundary of the bistable region coincides precisely with the LD-HD boundary.  In this case, as the initiation rate $\alpha$ passes through the LD-HD boundary, a discontinuity in the number of proteins occurs (Fig.~\ref{Fig5-crit-discont}A). [We obtain qualitatively the same behaviour by keeping $I$ fixed and varying  $k$.] This step-change in the number of proteins can be large, suggesting that small changes in the ribosome initiation rate $\alpha$ can result in a significant shift in protein levels.
Simulations confirm this  theoretical prediction. On increasing $\alpha$ dynamically during a simulation, a step change (around  75\% reduction) is clearly apparent on  transept  of the LD/HD critical value (Fig.~\ref{Fig5-crit-discont}B).  This cliff-edge response  is another unique feature of resulting from the interplay between feedback and recycling.

\noindent
{\bf Conclusions.}--Our model suggests  that negative and positive feedback acting together in translation may provide cells with a versatile mechanism to adapt their protein levels  according to the environment. Moreover, as mentioned above,  the centrally important protein PABP is subject to ribosome recycling and is known to  exhibit  translational negative feedback. Hence, understanding how its production is controlled is important to gain insight into translational control at the global level. Interestingly, PABP has  also been implicated in circadian oscillations~\cite{Kojima12}. The oscillatory behaviour  predicted  by our model  could therefore play an important role in  this fundamental  regulatory mechanism. Finally, disturbances of poly(A) tail length have been linked to a number of physiological and pathological processes. Therefore,  a better understanding of the interplay of ribosome recycling and translational negative feedback  has far reaching consequences.

\providecommand{\noopsort}[1]{}\providecommand{\singleletter}[1]{#1}%

\end{document}


\title{Translational control of gene expression via interacting feedback loops. Supplemental Material.}

\author{
Liang Wang\\ University of Dundee, Dundee,
United Kingdom
\and
M. Carmen Romano\\ University of Aberdeen, Aberdeen,
United Kingdom
\and
Fordyce A. Davidson*\\ University of Dundee, Dundee,
United Kingdom  \\*Corresponding Author
}

\maketitle
 
\section{Protein Dynamics and Effective Vacancy of the Start Site}\label{protein}
Protein complex formation and loss,  respectively, can be represented as follows:
\begin{displaymath}
\underbrace{\text{P}+\text{P}+\ldots\text{P}}_{n}\autorightleftharpoons{$r_{1}$}{$r_{2}$} n\text{P}, \quad \text{P} \autorightarrow{$r$}{} \emptyset
\end{displaymath}
where P is protein, $ n \geq 1 $ proteins form a protein complex $ n\text{P} $ and $ r_1,\,r_2 $ and $r$  are rate constants, where $r_1$ has dimensions of $(\text{volume})^{n-1}/\text{time}$, and $r_2$ and $r$ have dimensions of   1/time. 
Applying  a quasi-steady-state approximation for the complex concentration:
\begin{equation}
\dfrac{d[n\text{P}]}{dt}=r_1\left([\text{P}]\right)^{n}-r_2[n\text{P}]\approx 0\quad\Longrightarrow\quad [n\text{P}]=\dfrac{r_1}{r_2}\left([\text{P}]\right)^{n}=\dfrac{r_1}{r_2}\left(\dfrac{N}{V}\right)^n,\label{nP}
\end{equation}
where $ [\,\cdot\,] $ represents concentration, $ N $ denotes the number of molecules of P, and $ V $ denotes the cell volume.  The protein complex is assumed to  bind and unbind from some location at the start site (SS) at rates $k_1,  k_2 $, respectively:


\begin{equation} 
n\text{P}+\text{SS}\autorightleftharpoons{$k_{1}$}{$k_{2}$}n\text{P}\cdot\text{SS}\label{eqapp1}.
\end{equation}


Writing \eqref{eqapp1} as a differential equation and applying the quasi-steady state approximation as above  yields
\begin{align}
\cfrac{d[n\text{P}\cdot\text{SS}]}{dt}&= k_{1}[n\text{P}][\text{SS}]-k_{2} [n\text{P}\cdot\text{SS}]\approx 0 \nonumber\\
i.e.& \quad [n\text{P}\cdot\text{SS}] = \frac{k_1}{k_2} [n\text{P}] [\text{SS}].
\label{qss}
\end{align}

This quasi-steady state approximation leads to an {\it effective vacancy} of the SS from the perspective of ribosome binding events.  The probability that the  SS is vacant is equivalent to 
\begin{equation}\label{prob_free}
\frac{[\text{SS}]}{[\text{SS}] + [n\text{P}\cdot\text{SS}]} = \frac{[\text{SS}]}{[\text{SS}] + \frac{k_1}{k_2} [n\text{P}] [\text{SS}]}= \frac{k_2}{k_2+k_1 [n\text{P}]},
\end{equation}
where we have used \eqref{qss}.
This probability can be written in terms  the number of proteins, $N$, as follows. Using \eqref{nP} and \eqref{prob_free} it follows that the probability that the SS is free is 
\begin{equation}\label{feq}
f(N):=\dfrac{1}{1+\left(N/\theta\right)^{n}},
\end{equation}
where $ \theta=V\sqrt[n]{r_{2}k_2/r_{1}k_1} $.  


\section{Phases and Phase  Boundaries}
%
%
%

\noindent
{\bf Maximal Current Phase}: $ \alpha_{eff},\,\beta_{eff} \geqslant1/2 $  
 \begin{equation}
\rho_1=1-\dfrac{1}{4\alpha_{eff}},\qquad\rho_L=\dfrac{1}{4\beta_{eff}}, \qquad  \rho = \frac{1}{2}, \qquad J = \dfrac{1}{4}.\label{mcc}
\end{equation}

\noindent
{\bf Low Density Phase}: $ \alpha_{eff} <\beta_{eff},\,\alpha_{eff}<1/2 $
\begin{equation}
\rho_1=\alpha_{eff},\qquad\rho_L=\dfrac{\alpha_{eff}(1-\alpha_{eff})}{\beta_{eff}},\qquad  \rho = \alpha_{eff}, \qquad J = \alpha_{eff}(1-\alpha_{eff}). \label{ldc}
\end{equation}

\noindent
{\bf High Density Phase}: $ \alpha_{eff} >\beta_{eff},\,\alpha_{eff} > 1/2 $
\begin{equation}
\rho_1=1-\dfrac{\beta_{eff}(1-\beta_{eff})}{\alpha_{eff}},\qquad\rho_L=1-\beta_{eff},\qquad  \rho = 1-\beta_{eff}, \qquad J= \beta_{eff}(1-\beta_{eff}). \label{hdc}
\end{equation}
In each case the corresponding steady state protein level is given by $ N^*:= J/r$.  We now define the phase boundaries in a series of lemmas. Proofs are lengthy and hence grouped in the next section for  easier exposition.

\begin{lemma}
The Maximal Current Phase is defined by
\begin{equation}
\dfrac{\alpha}{2\alpha+k\left(1+I^n\right)}\leqslant\left\lbrace\begin{aligned}
&\beta&\leqslant\dfrac{\alpha k}{1+I^n-2\alpha}\quad&\text{if}\quad\alpha<\dfrac{1}{2}\left(1+I^n\right),\\
&\beta&\quad&\text{otherwise},
\end{aligned}\right.\label{pdformcI}
\end{equation}
where  $I:=\frac{1}{4r\theta}$. Within this region, 
\begin{small}
\begin{equation}
\alpha_{eff}=\dfrac{\alpha}{1+I^n}+\dfrac{k}{2\left(\beta+\sqrt{\beta\left(\beta+\dfrac{k}{\alpha}\left(1+I^n\right)\right)}\right)} \quad \mbox{and} \quad \beta_{eff}=\dfrac{1}{2}\left(\beta+\sqrt{\beta\left(\beta+\dfrac{k}{\alpha}\left(1+I^n \right)\right)}\right).
\end{equation}
\end{small}
\end{lemma}

\begin{remark}
Note that the stacking of the left and right inequalities \eqref{pdformcI} in  case $\alpha<\dfrac{1}{2}\left(1+I^n\right)$ is necessary and therefore generates a lower bound for $\alpha$ and $\beta$ marked by the intersection of the two functions of $\alpha$. It is straightforward to compute this intersection: 
\begin{equation}\label{intersect}
(\alpha, \beta) = \left( \frac{1-k}{2}\left(1+I^n\right),  \frac{1-k}{2}\right).
\end{equation}
\end{remark}


\begin{lemma}
The Low Density  Phase is defined by
\begin{equation}
\alpha<\left\lbrace\begin{aligned}
&\beta\left(1+\left(\dfrac{4 I (\beta+k)(1-\beta)}{(1+k)^2}\right)^n\right)&\qquad&\text{if}\quad 2\beta+k<1,\\
&\dfrac{\beta}{2\beta+k}\left(1+I^{n}\right)&\qquad&\text{otherwise}.
\end{aligned}\right.\label{pdforld}
\end{equation}
Within this region, there exists a unique, positive expression for $\alpha_{eff}$ that yields unique, positive expressions for $\beta_{eff}$ and $N^*$.
\end{lemma}


\begin{lemma}
The High Density  Phase  is defined by
\begin{equation}
\alpha>\left\lbrace\begin{aligned}
&\beta\left(1+\left(\dfrac{4 I(\beta+k)(1-\beta)}{(1+k)^2}\right)^n\right)&\qquad&\text{if}\quad 2\beta+k<1,\\
&\dfrac{k\beta}{1-2\beta}\left(1+I^n\right)&\qquad&\text{otherwise},
\end{aligned}\qquad\beta<\dfrac{1}{2}.\right.\label{pdforhdI}
\end{equation}
Within this region, there  exist either one or three positive solutions,  $\beta_{eff}$, that yield positive expressions for $\alpha_{eff}$ and $N^*$. 
\end{lemma}

\section{Proofs}
\noindent
{\it Proof of Lemma 1}
Substituting the effective rates  into \eqref{mcc}  yields:
\begin{align*}
\beta_{eff}&=\beta+k(1-\rho_1)=\beta+k\left(1-\left(1-\dfrac{1}{4\alpha_{eff}}\right)\right)=\beta+\dfrac{k}{4}\dfrac{1}{\alpha_{eff}}.\\
\end{align*}
Therefore, we obtain a quadratic equation for $ \beta_{eff} $:
\begin{displaymath}
4^{n+1}\alpha(\theta r)^n\beta_{eff}^2-4^{n+1}\alpha\beta(\theta r)^n\beta_{eff}-k\beta\left((4\theta r)^n+1\right)=0.
\end{displaymath}
Subsequently,\begin{align*}
\beta_{eff}=
&\dfrac{1}{2}\left(\beta\pm\sqrt{\beta^2+\dfrac{k\beta\left((4\theta r)^n+1\right)}{4^n\alpha(\theta r)^n}}\right).
\end{align*}
Since we require $ \beta_{eff} $ to be positive and $ \sqrt{\beta^2+\dfrac{k\beta\left((4\theta r)^n+1\right)}{4^n\alpha(\theta r)^n}}>\beta $, it follows that:\begin{small}
\begin{equation}
\beta_{eff}=\dfrac{1}{2}\left(\beta+\sqrt{\beta\left(\beta+\dfrac{k}{\alpha}\left(1+\left(\dfrac{1}{4\theta r}\right)^n\right)\right)}\right).
\end{equation}
\end{small}
Then $ \alpha_{eff} $ is easy to obtain:
\begin{align}
\alpha_{eff}
&=\dfrac{\alpha}{1+\left(\dfrac{1}{4\theta r}\right)^n}+\dfrac{k}{2\left(\beta+\sqrt{\beta\left(\beta+\dfrac{k}{\alpha}\left(1+\left(\dfrac{1}{4\theta r}\right)^n\right)\right)}\right)}. \label{aeffMC}
\end{align}
The maximal current phase is determined by the conditions $ \alpha_{eff},\,\beta_{eff}\geqslant\dfrac{1}{2} $ i.e.:\begin{displaymath}
\beta_{eff}=\dfrac{\beta+\sqrt{\beta\left(\beta+\dfrac{k}{\alpha}\left(1+\left(\dfrac{1}{4\theta r}\right)^n\right)\right)}}{2}\geqslant\dfrac{1}{2},
\end{displaymath}
so $ \sqrt{\beta\left(\beta+\dfrac{k}{\alpha}\left(1+\left(\dfrac{1}{4\theta r}\right)^n\right)\right)}\geqslant1-\beta $. Since $ 1-\beta>0 $ for $ \beta<1 $, we have
\begin{equation}
\beta\geqslant\dfrac{\alpha}{2\alpha+k\left(1+\left(\dfrac{1}{4\theta r}\right)^n\right)},
\end{equation}
and
\begin{displaymath}
\alpha_{eff}=\dfrac{\alpha}{1+\left(\dfrac{1}{4\theta r}\right)^n}+\dfrac{k}{2\left(\beta+\sqrt{\beta\left(\beta+\dfrac{k}{\alpha}\left(1+\left(\dfrac{1}{4\theta r}\right)^n\right)\right)}\right)}\geqslant\dfrac{1}{2},
\end{displaymath}
namely,\begin{equation}
\dfrac{k}{2\left(\beta+\sqrt{\beta\left(\beta+\dfrac{k}{\alpha}\left(1+\left(\dfrac{1}{4\theta r}\right)^n\right)\right)}\right)}\geqslant\dfrac{1}{2}-\dfrac{\alpha}{1+\left(\dfrac{1}{4\theta r}\right)^n}.\label{mccondition}
\end{equation}
If $ \dfrac{1}{2}-\dfrac{\alpha}{1+\left(\dfrac{1}{4\theta r}\right)^n}\leqslant 0 $, i.e., $ \alpha\geqslant\dfrac{1}{2}\left(1+\left(\dfrac{1}{4\theta r}\right)^n\right) $, then \eqref{mccondition} is always true. 
\par
If $ \alpha<\dfrac{1}{2}\left(1+\left(\dfrac{1}{4\theta r}\right)^n\right) $, we have:\begin{displaymath}
k\geqslant2\left(\dfrac{1}{2}-\dfrac{\alpha}{1+\left(\dfrac{1}{4\theta r}\right)^n}\right)\left(\beta+\sqrt{\beta\left(\beta+\dfrac{k}{\alpha}\left(1+\left(\dfrac{1}{4\theta r}\right)^n\right)\right)}\right),
\end{displaymath}
subsequently,\begin{small}
\begin{align}
k-\left(1-\dfrac{2\alpha}{1+\left(\dfrac{1}{4\theta r}\right)^n}\right)\beta\geqslant\left(1-\dfrac{2\alpha}{1+\left(\dfrac{1}{4\theta r}\right)^n}\right)\sqrt{\beta\left(\beta+\dfrac{k}{\alpha}\left(1+\left(\dfrac{1}{4\theta r}\right)^n\right)\right)}>0.\label{mccondition1}
\end{align}
\end{small}
\par
From \eqref{mccondition1}, we have a weak condition

%
\begin{equation}
\beta\leqslant\dfrac{\alpha k}{1+\left(\dfrac{1}{4\theta r}\right)^n-2\alpha}.\label{mcbeta2}
\end{equation}
The result follows by grouping the inequalities determined above and on writing the final expression in terms of the intensity factor.

\hfill $\Box$
\vspace{0.5cm}

\noindent
{\it Proof of Lemma 2}
Substituting  the effective rates  into \eqref{ldc} yields:
\begin{align*}
\alpha_{eff}=
\dfrac{\alpha}{1+\left(\dfrac{\alpha_{eff}(1-\alpha_{eff})}{\theta r}\right)^n}+k\dfrac{\alpha_{eff}(1-\alpha_{eff})}{\beta+k(1-\alpha_{eff})}.
\end{align*}
Again after some manipulation we obtain:
\begin{equation}
P(\alpha_{eff}):= \beta\alpha_{eff}^{n+1}(1-\alpha_{eff})^n+(\theta r)^n(\beta+\alpha k)\alpha_{eff}-\alpha(\theta r)^n(\beta+k)=0.\label{P}
\end{equation}
If  there exist positive solutions, $\alpha_{eff}$, to \eqref{P}, then  from \eqref{ldc}, $ \beta_{eff}=\beta+k\left(1-\alpha_{eff}\right) $ and $N^* = \alpha_{eff}(1-\alpha_{eff})/r$. 
Next we consider the conditions for the existence of $ \alpha_{eff} $ and thus define  the phase boundary for the LD phase.  LD  phase is by  defined by $ \alpha_{eff}<\beta_{eff},\,\alpha_{eff}<\dfrac{1}{2} $. From $ \alpha_{eff}<\beta_{eff} $, we have:\begin{align*}
\alpha_{eff}<\beta_{eff}=\beta+k(1-\alpha_{eff}),\quad\Longrightarrow\quad (1+k)\alpha_{eff}<\beta+k,
\end{align*}
therefore\begin{equation}
\alpha_{eff}<\dfrac{\beta+k}{1+k}.
\end{equation}
On the other hand, we require $ \alpha_{eff}<\dfrac{1}{2} $. Therefore $ \alpha_{eff}<\min\left(\dfrac{1}{2},\,\dfrac{\beta+k}{1+k}\right) $.
\par
\par
If $ \dfrac{\beta+k}{1+k}\geqslant\dfrac{1}{2} $, i.e., $ 2\beta+2k\geqslant1+k $, $ 2\beta+k\geqslant1 $, then $ \alpha_{eff}<\dfrac{1}{2} $. The derivative of $P$ with respect to $\alpha_{eff} $ is\begin{align*}
P^\prime
&=\beta\alpha_{eff}^{n}(1-\alpha_{eff})^{n-1}\left(1-\alpha_{eff}+n(1-2\alpha_{eff})\right)+(\theta r)^n(\beta+\alpha k).
\end{align*}
Since both $ 1-\alpha_{eff} $ and $ 1-2\alpha_{eff} $ are positive in $ \left(0,\,\dfrac{1}{2}\right) $, so $ P^\prime>0 $ in $ \left(0,\,\dfrac{1}{2}\right) $.  Moreover, $ P(0)=-\alpha(\theta r)^n(\beta+k)<0 $. Therefore, a real, positive solution of \eqref{P} exists if and only if $ f\left(\dfrac{1}{2}\right)>0 $ (in this case, the solution is unique). Now,  
\begin{align*}
P\left(\dfrac{1}{2}\right)&=
\dfrac{(\theta r)^n}{2}\left(\beta\left(1+\left(\dfrac{1}{4\theta r}\right)^{n}\right)-\alpha(2\beta+k)\right),
\end{align*}
so $ P\left(\dfrac{1}{2}\right)>0 $ requires $ \beta\left(1+\left(\dfrac{1}{4\theta r}\right)^{n}\right)-\alpha(2\beta+k)>0 $, namely,\begin{equation}
\alpha<\dfrac{\beta}{2\beta+k}\left(1+\left(\dfrac{1}{4\theta r}\right)^{n}\right).\label{ldcondition}
\end{equation} 
If $ \dfrac{\beta+k}{1+k}<\dfrac{1}{2} $, i.e., $ 2\beta+k<1 $, then $ \alpha_{eff}<\dfrac{\beta+k}{1+k} $. Since $ P $ is increasing in $ \left(0,\,\dfrac{1}{2}\right) $ and $ P(0)<0 $, so a solution of \eqref{P} exists if and only if $ P\left(\dfrac{\beta+k}{1+k}\right)>0 $ (and again the solution is unique).
\begin{align*}
P\left(\dfrac{\beta+k}{1+k}\right)&=
\dfrac{(\beta+k)(\theta r)^n}{1+k}\left(\beta\left(\dfrac{(\beta+k)(1-\beta)}{\theta r(1+k)^2}\right)^n+\beta-\alpha\right)>0,
\end{align*}
provided
\begin{equation}
\alpha<\beta\left(1+\left(\dfrac{(\beta+k)(1-\beta)}{\theta r(1+k)^2}\right)^n\right).\label{ldcondition1}
\end{equation}
The proof is complete by writing the inequalities in terms of the Intensity Factor.

\hfill $\Box$
\vspace{0.5cm}

\noindent
{\it Proof of Lemma 3}  Substituting  the effective rates into \eqref{hdc}  yields:

\begin{align*}
\beta_{eff}
&=\beta+k\dfrac{\beta_{eff}(1-\beta_{eff})}{\dfrac{\alpha}{1+\left(\dfrac{N}{\theta}\right)^n}+k\rho_{L}}=\beta+\dfrac{k\beta_{eff}(1-\beta_{eff})}{\dfrac{\alpha}{1+\left(\dfrac{\beta_{eff}(1-\beta_{eff})}{\theta r}\right)^n}+k(1-\beta_{eff})}.
\end{align*}
Further algebra yields the equation
\begin{equation}
Q(\beta_{eff}):= k\beta\beta_{eff}^n\left(1-\beta_{eff}\right)^{n+1}-(k\beta+\alpha)(\theta r)^n\beta_{eff}+\beta(\theta r)^n(\alpha+k)=0.\label{Q}
\end{equation}
With $ \beta_{eff} $  a solution of \eqref{Q}, then from \eqref{hdc},   $\rho=1-\beta_{eff} $, $ J=\beta_{eff}(1-\beta_{eff}) $ and $N^* = \beta_{eff}(1-\beta_{eff})/r$. Hence, the effective initiation rate is given by 
\[
 \alpha_{eff}=\dfrac{\alpha}{1+\left(\dfrac{\beta_{eff}(1-\beta_{eff})}{\theta r}\right)^n}+k\left(1-\beta_{eff}\right).
 \]

%
One of the HD conditions is
 $ \beta_{eff}<\dfrac{1}{2} $,  i.e.  $ \dfrac{1}{2}>\beta_{eff}=\beta+k(1-\rho_1)>\beta $, namely, $ \beta<\dfrac{1}{2} $.
Moreover, from the second condition for HD $ \alpha_{eff}>\beta_{eff} $, we have\begin{displaymath}
\beta_{eff}=\beta+k\dfrac{\beta_{eff}(1-\beta_{eff})}{\alpha_{eff}}<\beta+k(1-\beta_{eff})\quad\Longrightarrow\quad (1+k)\beta_{eff}<\beta+k.
\end{displaymath}
Therefore, we require $ \beta_{eff}<\min\left\{\dfrac{1}{2},\,\dfrac{\beta+k}{1+k}\right\}$.
\par
If $ \dfrac{\beta+k}{1+k}\geqslant\dfrac{1}{2} $, i.e., $ 2\beta+k\geqslant1 $ then we  need to check the existence of solutions to \eqref{Q} in $ \left(0,\,\dfrac{1}{2}\right) $. To this end, note that 
\begin{align*}
Q^\prime
&=k\beta\beta_{eff}^{n-1}\left(1-\beta_{eff}\right)^{n}\left(n-(2n+1)\beta_{eff}\right)-(k\beta+\alpha)(\theta r)^n,\\
Q^{\prime\prime}
&=nk\beta\beta_{eff}^{n-2}\left(1-\beta_{eff}\right)^{n-1}\left(2(2n+1)\beta_{eff}^2-4n\beta_{eff}+n-1\right).
\end{align*}
Thus, $ Q^{\prime\prime}>0 $  in $ \left(0,\,\beta_{eff}^-\right) $ and $ Q^{\prime\prime}<0 $  in $ \left[\beta_{eff}^-,\,\dfrac{1}{2}\right) $ with the maximum of $Q'$ obtained at  $\beta_{eff}^-$. Moreover,
\begin{align*}
Q^\prime\left(0\right)&=-(k\beta+\alpha)(\theta r)^n<0\quad \mbox{and}\quad
Q^\prime\left(\dfrac{1}{2}\right)=-k\beta\dfrac{1}{4^n}-(k\beta+\alpha)(\theta r)^n<0.
\end{align*}
If $ Q^\prime\left(\beta_{eff}^-\right)<0 $, then $ Q^\prime<0 $ ($ Q $ is decreasing) in $ \left(0,\,\dfrac{1}{2}\right) $. Besides, $ Q(0)=\beta(\theta r)^n(\alpha+k)>0 $. Thus, the solution of \eqref{Q} $ Q=0 $ exists if and only if $ Q\left(\dfrac{1}{2}\right)<0 $, and the solution is unique.
\par
If $ Q^\prime\left(\beta_{eff}^-\right)>0 $, then there are two solutions to $ Q^\prime=0 $. Let $ \beta_{eff}^1 $ and $ \beta_{eff}^2 $ denote  the solutions and $ 0<\beta_{eff}^1<\beta_{eff}^2<\dfrac{1}{2} $. Thus, $ Q^\prime<0 $ ($ Q $ is decreasing) in $ \left(0,\,\beta_{eff}^1\right)\cup\left(\beta_{eff}^2,\,\dfrac{1}{2}\right) $, $ Q^\prime>0 $ ($ Q $ is increasing) in $ \left(\beta_{eff}^1,\,\beta_{eff}^2\right) $. Moreover, $ Q(0)=\beta(\theta r)^n(\alpha+k)>0 $, and $ Q\left(\beta_{eff}^1\right) $ is the local minimum. We have following cases. 
\par
(I) If $ Q\left(\beta_{eff}^1\right)<0 $ and $ Q\left(\beta_{eff}^2\right)<0 $, then $ Q<0 $ in $ \left(\beta_{eff}^1,\,\dfrac{1}{2}\right) $, there exists a unique solution to \eqref{Q} in $ \left(0,\,\beta_{eff}^1\right) $. In this case, $ Q\left(\dfrac{1}{2}\right)<0 $. \par
(III) If $ Q\left(\beta_{eff}^1\right)<0 $ and $ Q\left(\beta_{eff}^2\right)>0 $, then \eqref{Q}  has three solutions if $ Q\left(\dfrac{1}{2}\right)<0 $ or two solutions if $ Q\left(\dfrac{1}{2}\right)>0 $. If $ Q\left(\dfrac{1}{2}\right)>0 $, we have:
\begin{align*}
Q\left(\dfrac{1}{2}\right)
&=\dfrac{(\theta r)^n}{2}\left(k\beta\left(\dfrac{1}{4\theta r}\right)^n-\alpha+2\beta\alpha+\beta k\right)>0,
\end{align*}
which gives:
\begin{equation}\label{contra}
\beta>\dfrac{\alpha}{2\alpha+k\left(1+\left(\dfrac{1}{4\theta r}\right)^n\right)}.
\end{equation}
However, from \eqref{pdformcI}, \eqref{contra} sets the system in the MC phase and hence is not a feasible condition.  Therefore, in this case, $ Q\left(\dfrac{1}{2}\right)<0 $ and there are three solutions to \eqref{Q}.\par
Overall, for $ 2\beta+k\geqslant1 $, the existence of the solution of $ \beta_{eff} $ requires $ Q\left(\dfrac{1}{2}\right)<0 $:
\begin{align*}
Q\left(\dfrac{1}{2}\right)
&=\dfrac{(\theta r)^n}{2}\left(k\beta\left(\dfrac{1}{4\theta r}\right)^n-\alpha+2\beta\alpha+\beta k\right)<0,
\end{align*}
which gives:\begin{equation}
\alpha>\dfrac{k\beta\left(1+\left(\dfrac{1}{4\theta r}\right)^n\right)}{1-2\beta}.\label{hdcondition}
\end{equation}

\par
If $ \dfrac{\beta+k}{1+k}<\dfrac{1}{2} $, i.e., $ 2\beta+k<1 $, then  we require $ \beta_{eff}<\dfrac{\beta+k}{1+k} $. Similar to the case for  $ 2\beta+k\geqslant1 $, the existence of the solution of $ \beta_{eff} $ requires $ Q\left(\dfrac{\beta+k}{1+k}\right)<0 $ and one or three solutions can be obtained. Note, 
\begin{align*}
Q\left(\dfrac{\beta+k}{1+k}\right) 
&=\dfrac{(\theta r)^nk(1-\beta)}{1+k}\left(\beta\left(\dfrac{(\beta+k)(1-\beta)}{\theta r(1+k)^2}\right)^n+\beta-\alpha\right)<0,
\end{align*}
provided 
\begin{equation}
\alpha>\beta\left(1+\left(\dfrac{(\beta+k)(1-\beta)}{\theta r(1+k)^2}\right)^n\right).
\end{equation}
The result follows directly as above.

\hfill $\Box$
\vspace{0.5cm}

\section{Simulation}
In every Monte-Carlo step (MCS), $L+2$ random numbers were chosen, corresponding to all possible reactions. 
Protein degradation was modelled using $n_r$ (the current number of proteins) attempts, each with  corresponding degradation rate, $r$. The initiation and termination  probabilities  were governed by $\alpha_{eff}$ and $\beta_{eff}$.  The processes of  complex formation and subsequent interaction with the Start Site were implemented using the quasi-steady state approximations detailed above. [Test cases where all processes were explicitly accounted, such as protein complex formation and protein complex binding, etc.,  displayed  little qualitative or quantitative difference to the hybrid Monte-Carlo used here,  but  required  significantly greater computational time.]

Time average was calculated from a simulation consisting of $10^8$ Monte-Carlo steps (MCS), where the first $10^6$ MCS were disregarded to ensure  steady-state. Then, the integration time was divided in windows of $10^4$ MCS, so that the standard deviation could be calculated for each value of $\alpha$. 

Typical simulations  are shown in Fig.S1A. The lag before the first protein is made  is approximately 500 time units corresponding to the length of the lattice.  After a transient rise in protein number, the  model predicts steady (average) protein number. The average long term behaviour of the Monte-Carlo simulations matched closely  the corresponding  analytical steady states $N^*$   derived above.  

%
In Fig. S1B, the phase diagram for the average  particle density is shown. Three separate  phases  are clearly identified by relatively sharp changes  in  average density. This phase separation is well-approximated by the analytical phase boundaries for the modified TASEP derived above. Within the phases the qualitative match of the phase density with that predicted by the steady state theory is generally in good agreement - this is easiest identified in the maximal current phase where the steady state theory predicts the value $\rho = 1/2$.  Agreement is generally good in the other phases.

\begin{figure}[h]
\begin{minipage}{0.5\linewidth}
\includegraphics[width=\textwidth]{../Figures/FigS1-MC-MI}
\end{minipage}
\begin{minipage}{0.45\linewidth}
\includegraphics[width=\textwidth]{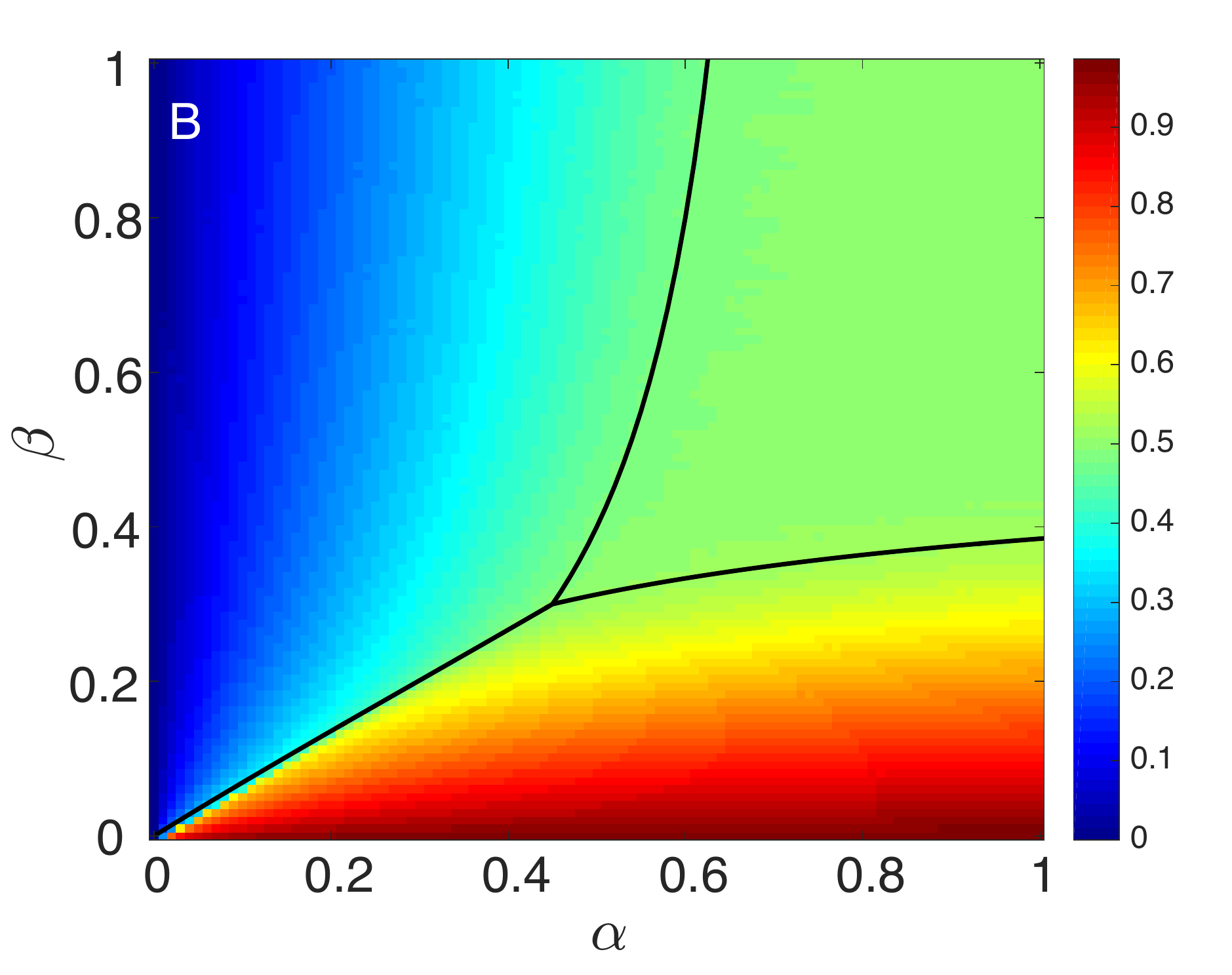}
\end{minipage}
\caption{Typical phase diagram and associated simulation. (A) Monte-Carlo simulations  as detailed in the text  showing  the temporal evolution of protein copy number. Averages over 5,000 realizations of stochastic simulations (black lines),  twenty example realizations (grey lines), single example trajectory (green line). Red dashed lines are mean field $ N^\ast $ as computed using the steady state theory. $L=500$ sites,  $r=0.005$, $I=0.05$, $k=0.4$, and $n=1$.  (B) Monte-Carlo simulations showing the average density of ribosomes,  $\rho$,  on the mRNA depending on $\alpha$ and $\beta$.  5000 simulations were used for each representative $(\alpha, \beta)$- pair.   The black lines show the analytical mean-field phase boundaries derived above.}\label{FigS2}
\end{figure}

%
%


\section{Dynamics in the Modified TASEP}

\subsection{Onset of oscillations is  characterised by a Hopf Bifurcation}
\begin{figure}[h]\label{Fig-Hopf}
\center
\includegraphics[width=0.5\textwidth]{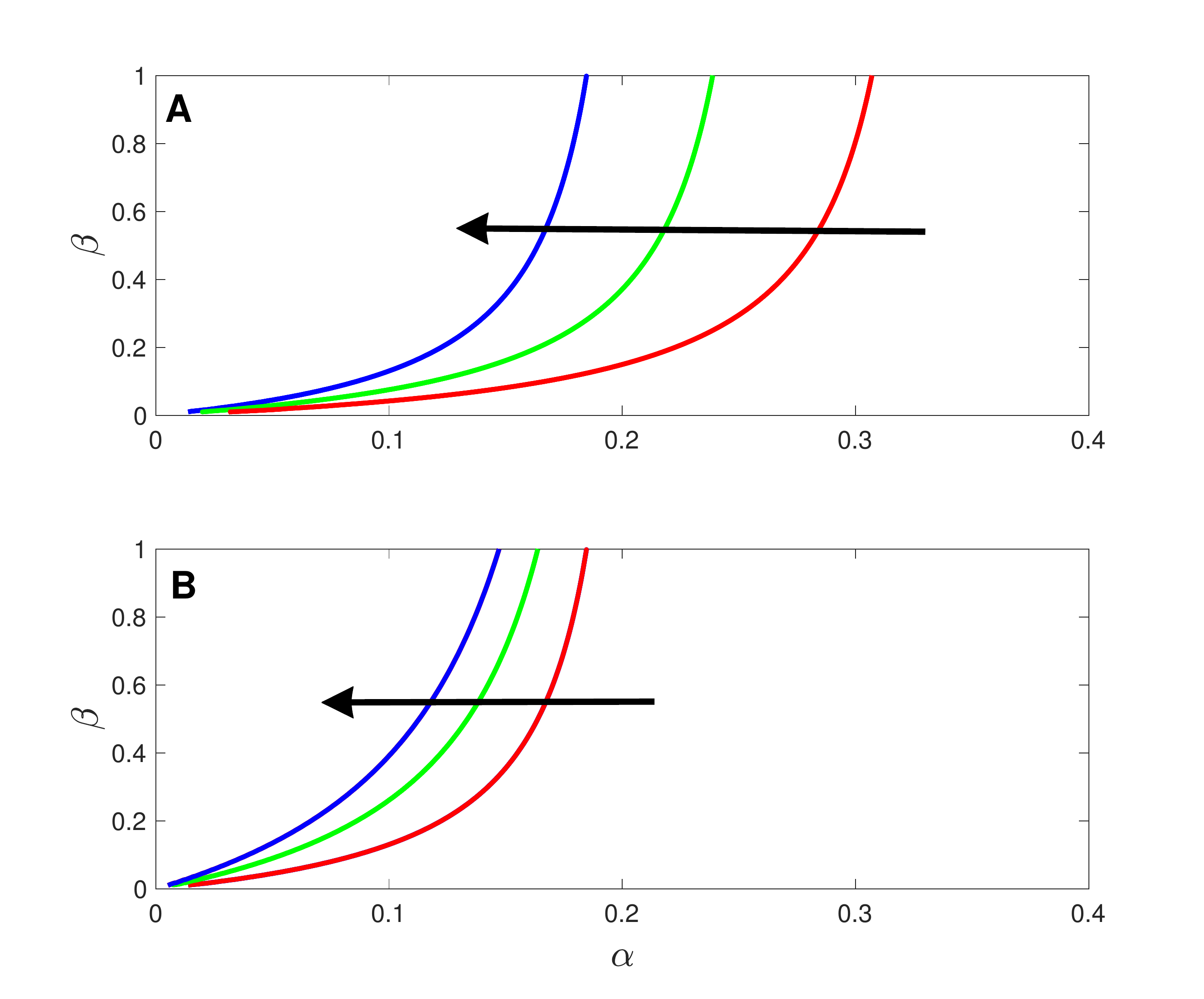}
\caption{The Hopf locus and the effect of varying recycling and feedback.  Hopf bifurcation locus in the $\alpha-\beta$ plane computed using \eqref{hopf}.  The effect of varying feedback intensity (A) and recycling (B) is shown.   Black arrows indicate direction of increasing effect. (A) $I=10/4$ (blue), $I=10/6$ (green), $I=25/9$ (red) with $k=0.2$ (B) $k=0.6$ (red),  $k=0.4$ (blue), $k=0.2$ (red) with $I = 10/4$. $L=500$ sites, $r=0.002$, $k=0.2$,  $n=5$.}
\end{figure}

The mean time for a particle to transit the lattice can be estimated as
\begin{equation}\label{meantransitT}
T = \frac{L}{1-\langle \rho \rangle} = \frac{L}{1-\langle \alpha_{eff} \rangle}
\end{equation}
%
in the LD regime. After substitution and some rearranging,  the effective ribosome binding rate in LD is 
\begin{equation} \label{ae}
\alpha_{eff}(N)=\dfrac{\alpha(\beta+k)}{\alpha k+\beta\left(1+\left(N/\theta\right)^n\right)}.
\end{equation}
Appealing to  the mean field approximation and setting $\langle N\rangle = N^*$,  the mean transit time is then defined as 
\[
\langle T\rangle= \frac{L}{1-\frac{\alpha(\beta+k)}{\alpha k +\beta \left(1+ \left(\frac{\langle N \rangle}{\theta} \right)^n\right)}}.
\]
The delay differential equation for protein copy number in the LD regime is: 
\begin{equation}
\dfrac{dN(t)}{dt}=\dfrac{\alpha(\beta+k)}{\alpha k+\beta\left(1+\left(\dfrac{N(t-T)}{\theta}\right)^n\right)}\left(1-\dfrac{\alpha(\beta+k)}{\alpha k+\beta\left(1+\left(\dfrac{N(t-T)}{\theta}\right)^n\right)}\right)-rN(t),\label{DD1}
\end{equation}
 where we have written $T$ for $\langle T\rangle$ for ease of exposition here and below. Steady states, $ N^\ast $, of \eqref{DD1} are  exactly the solutions generated by substituting  solutions of \eqref{P} into the definition $N^* = \alpha_{eff}(1-  \alpha_{eff})/r$. 
To check the stability of $ N^\ast $, we consider perturbations, $\delta(t)$, with   $ \vert\delta \vert\ll 1 $. Substituting $ N(t)=N^{\ast}+\delta(t) $ into \eqref{DD1}, using Taylor expansion and   dropping the higher order term yields \begin{equation}
\dfrac{d\delta(t)}{dt}=-B(N^*)\delta(t-T)-r\delta(t), \label{DDp}
\end{equation}
where 
 \begin{equation}
B(N^*):=\left(1-\dfrac{2\alpha(\beta+k)}{\alpha k+\beta\left(1+\left(\dfrac{N}{\theta}\right)^n\right)}\right)\dfrac{\alpha\beta(\beta+k)n\left(\dfrac{N}{\theta}\right)^n}{N\left(\alpha k+\beta\left(1+\left(\dfrac{N}{\theta}\right)^n\right)\right)^2}>0\label{DDb}
\end{equation}
as $\alpha_{eff} < 1/2$  in the LD phase by definition. Setting $\delta(t)=Ae^{\lambda t} $ and  on substitution  into \eqref{DDp}, we get (hereon setting $B(N^*) = B$ for ease of notation) 
\begin{equation}
\lambda+r=-Be^{-\lambda T}.\label{eigenDD}
\end{equation}
Setting  $ \lambda =\mu + i \omega $, standard arguments yield a critical wave number at which stability is lost, namely,  
 $\omega^2 = B^2-r^2$ and hence a necessary condition is  $B>r$.  Some algebra  eliminates $\omega$ to defines the Hopf locus in terms of system parameters thus:
\begin{equation}\label{hopf}
B \cos\left(\sqrt{B^2-r^2} \, T\right) +r =0.
\end{equation}
Hence, the existence of periodic solutions to \eqref{DD1} is guaranteed, at least for values of $(\alpha,\beta)$ sufficiently close to the locus.  The Hopf locus in the $\alpha-\beta$ plane is shown in Figure~S\ref{Fig-Auto-Corr} along with   the effects of varying $I$ and $k$ (Figs.~S3 A and B).
Finally, note that the  necessary condition $ B-r >0$, yields: 
\[
\dfrac{\alpha\beta(\beta+k)}{\left(\alpha k+\beta\left(1+\left(\dfrac{N^\ast}{\theta}\right)^n\right)\right)^3}g\left(\left(\dfrac{N^\ast}{\theta}\right)^n\right) >0,
\]
where
\begin{align}
g\left(\left(\dfrac{N^\ast}{\theta}\right)^n\right)=&\beta(n-1)\left(\dfrac{N^\ast}{\theta}\right)^{2n}-(\beta+\alpha k)(1-\alpha)\notag\\
&+\left(\beta(\alpha+n-2\alpha n-2)-(n+1)\alpha k\right)\left(\dfrac{N^\ast}{\theta}\right)^n.
\end{align}
Hence, $B-r>0 \implies  g\left(\left(\dfrac{N^\ast}{\theta}\right)^{n}\right) >0$. Notice that if $ n=1 $, $ g\left(\left(\dfrac{N^\ast}{\theta}\right)^{n}\right)=\left(-(\alpha+1)\beta-2\alpha k\right)\dfrac{N^\ast}{\theta}-(\beta+\alpha k)(1-\alpha)<0 $. Hence the Hopf bifurcation cannot occur in this case. Therefore, co-operativity is a necessary condition for Hopf bifurcation to occur. 
\par
For $ n>1 $, the discriminant of $ g\left(\left(\dfrac{N^\ast}{\theta}\right)^{n}\right) $ is given by:\begin{align} \label{discri}
\Delta=\left(\beta(\alpha+n-2\alpha n-2)-(n+1)\alpha k\right)^2+4\beta(n-1)(\beta+\alpha k)(1-\alpha),
\end{align}
and $\Delta >0$ certainly when $ \alpha<1 $. Therefore, in this case $ g\left(\left(\dfrac{N^\ast}{\theta}\right)^{n}\right)=0 $ has two distinct real roots:\begin{displaymath}
\left(\dfrac{N^\ast}{\theta}\right)^{n}_{\pm}=\dfrac{-\left(\beta(\alpha+n-2\alpha n-2)-(n+1)\alpha k\right)\pm\sqrt{\Delta}}{2\beta(n-1)},
\end{displaymath}
where $ \Delta $ is given in \eqref{discri}  and $ \left(\dfrac{N^\ast}{\theta}\right)^{n}_{-} $ is negative.  Hence, it is easy to conclude that $ g\left(\left(\dfrac{N^\ast}{\theta}\right)^{n}\right)>0 $ requires:\begin{equation}
\left(\dfrac{N^\ast}{\theta}\right)^{n}>\left(\dfrac{N^\ast}{\theta}\right)^{n}_{+}=\dfrac{-\left(\beta(\alpha+n-2\alpha n-2)-(n+1)\alpha k\right)+\sqrt{\Delta}}{2\beta(n-1)}.\label{DDs}
\end{equation}
The inequality \eqref{DDs} and setting the  right hand side of \eqref{DD1} to zero,  together imply 
\begin{equation}
\theta r<\dfrac{\alpha(\beta+k)}{\alpha k+\beta\left(1+\left(\dfrac{N^\ast}{\theta}\right)^n_{+}\right)}\left(1-\dfrac{\alpha(\beta+k)}{\alpha k+\beta\left(1+\left(\dfrac{N^\ast}{\theta}\right)^n_{+}\right)}\right)\dfrac{1}{\sqrt[n]{\left(\dfrac{N^\ast}{\theta}\right)^{n}_{+}}}.\label{DDbf}
\end{equation}
Rewriting this in terms of the Intensity Factor $I$ yields the condition $I> F(\bf{\Lambda})$
where $F(\bf{\Lambda})$ is a multiple of the  reciprocal of the right hand side of \eqref{DDbf}. 

\subsection{Numerical Simulations of the Characteristic DDE}
The DDE \eqref{DD1} was   solved numerically using the code   \texttt{Matlab dde}.
Typical behaviour for a range of values of $\alpha$ is shown in the Figures below. 
\begin{figure}[h]\label{FigS5}
\center
\includegraphics[width=\textwidth]{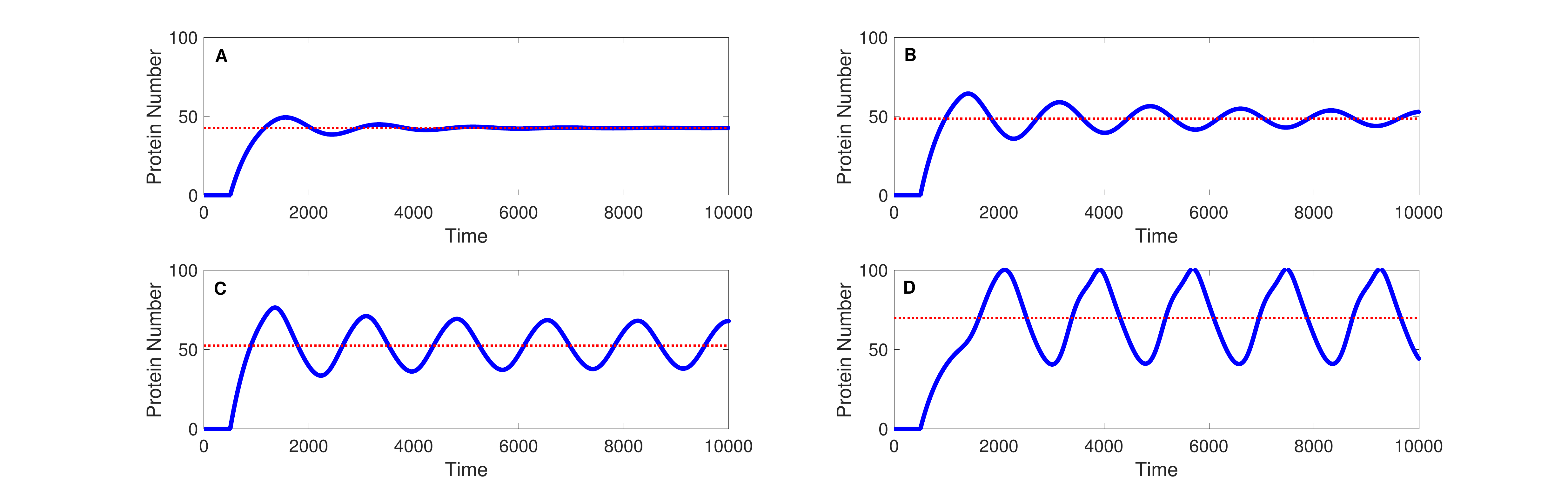}
\caption{Onset of periodic oscillations in DDE simulation. Solutions of DDE \eqref{DD1} showing onset of periodic solutions as $\alpha$ is increased through the Hopf locus shown in Figure S4.  Blue lines show the numerical solution of \eqref{DD1}.  The red dotted line shows the steady state value $N^*$ as predicted by the steady state theory.    (A) $\alpha = 0.1$, (B) $\alpha = 0.15$, (C) $\alpha = 0.2$, (D) $\alpha = 0.8$. Otherwise $L=500$ sites, $\theta=50$, $r=0.002$, $k=0.2$,  $n=5$,  $\beta = 0.5$.}
\end{figure}
The amplitude and period of the oscillations generated by \eqref{DD1} are in good agreement with those generated by Monte-Carlo simulations  (cf. Fig. S3D and Fig.2A in the main text). 

\subsection{Periodic oscillations are restricted to  the  LD Phase}
\begin{figure}[h]\label{Fig-Auto-Corr}
\center
\includegraphics[width=0.4\textwidth]{../Figures/hopf_locus_analy_num.pdf}
\caption{Periodicity in the Monte-Carlo simulations quantified by the autocorrelation function (heat map) compared to the  Hopf bifurcation locus in the $\alpha-\beta$ plane analytically calculated using \eqref{hopf} (upper black curve) and the LD/HD boundary from the mean-field approach (lower black curve).  Heatmap shows an autocorrelation-based measure to numerically detect oscillations in the time series of the number of proteins. In particular, we plot the value of the autocorrelation function $ACF(\tau)$ at the first maximum for $\tau>0$. For a maximum to be ``well-defined", the autocorrelation function must cross the horizontal axis twice. If there is no well-defined maximum, we set the measure equal to $0$. Hence, large values (red) in the heatmap indicate pronounced oscillations in the number of proteins. $I = 10/4$,  $L=500$ sites, $r=0.002$, $k=0.2$,  $n=5$.}
\end{figure}
In the MC phase, $J \equiv 1/4$ and $\alpha_{eff}$ is given by \eqref{aeffMC} and thus the loading rate is independent of time.  Therefore  $N\equiv 1/4r$ is the unique  asymptotic solution related to any non-zero initial data. Similarly,  in the HD phase,  $J = \beta_{eff} (1- \beta_{eff})$, with $\beta_{eff}$   given by the solutions of \eqref{Q}.  In this case, $\beta_{eff} = \beta_{eff}(N(t))$ and thus $J=J(N(t))$ and the characteristic equation $dN/dt = J-rN$  is a scalar, ordinary differential equation with constant coefficients.  Standard theory precludes periodic solutions.  The behaviour of the characteristic equations therefore supports the numerical simulations to suggest that (almost) periodic solutions are only realised in the LD phase. 
 
We computed an autocorrelation function based measure $ACF(\tau):=E[(N(t)-\mu)(N(t+\tau)-\mu)]/\sigma^2$,
where $\mu$ is the average and $\sigma$ the standard deviation of the time series $N(t)$ of the number of proteins. $N(t)$ means the time series at time $t$, and $N(t+\tau)$ the time series at time $t+\tau$. Results show that the Hopf locus associated with the DDE provides a reasonable approximation for the parameter region in the $\alpha-\beta$ plane for which oscillations are observed (Fig.~S4).

%
%
%

\newpage
\section{Bistability}
\begin{figure}[h]
\center
\includegraphics[width=0.5\textwidth]{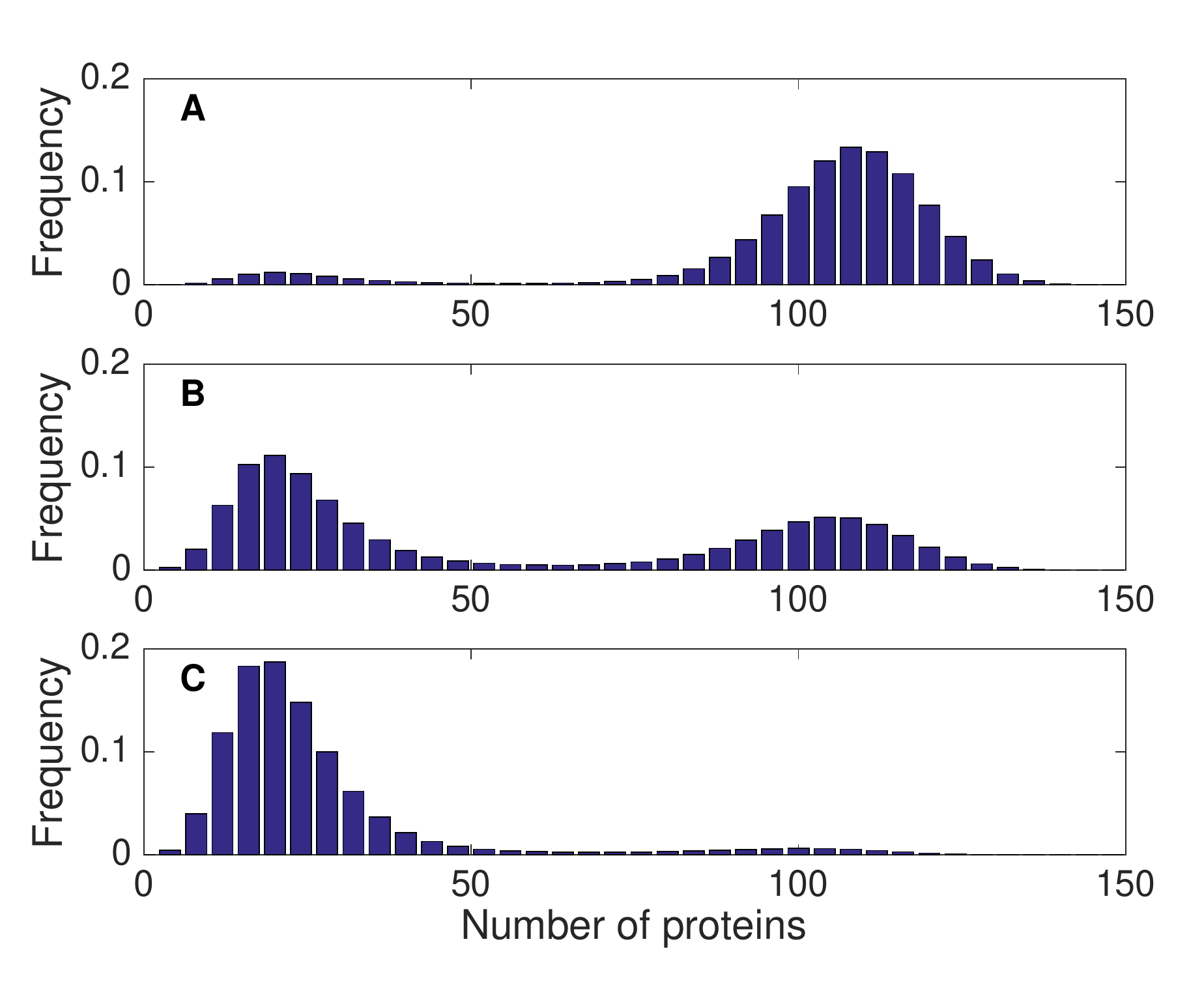}
\caption{Histograms of the time series of the number of proteins in the bistable state for $L=500$ sites, $\theta=21$, $d=0.002$, $k=0.8$, $\beta=0.015$ and $n=2$ and three different values of $\alpha$: (A) $\alpha=0.75$, (B) $\alpha=0.77$ and (C) $\alpha=0.79$.}\label{FigS_bistability_histograms}
\end{figure}

\begin{figure}[h]
\center
\includegraphics[width=0.5\textwidth]{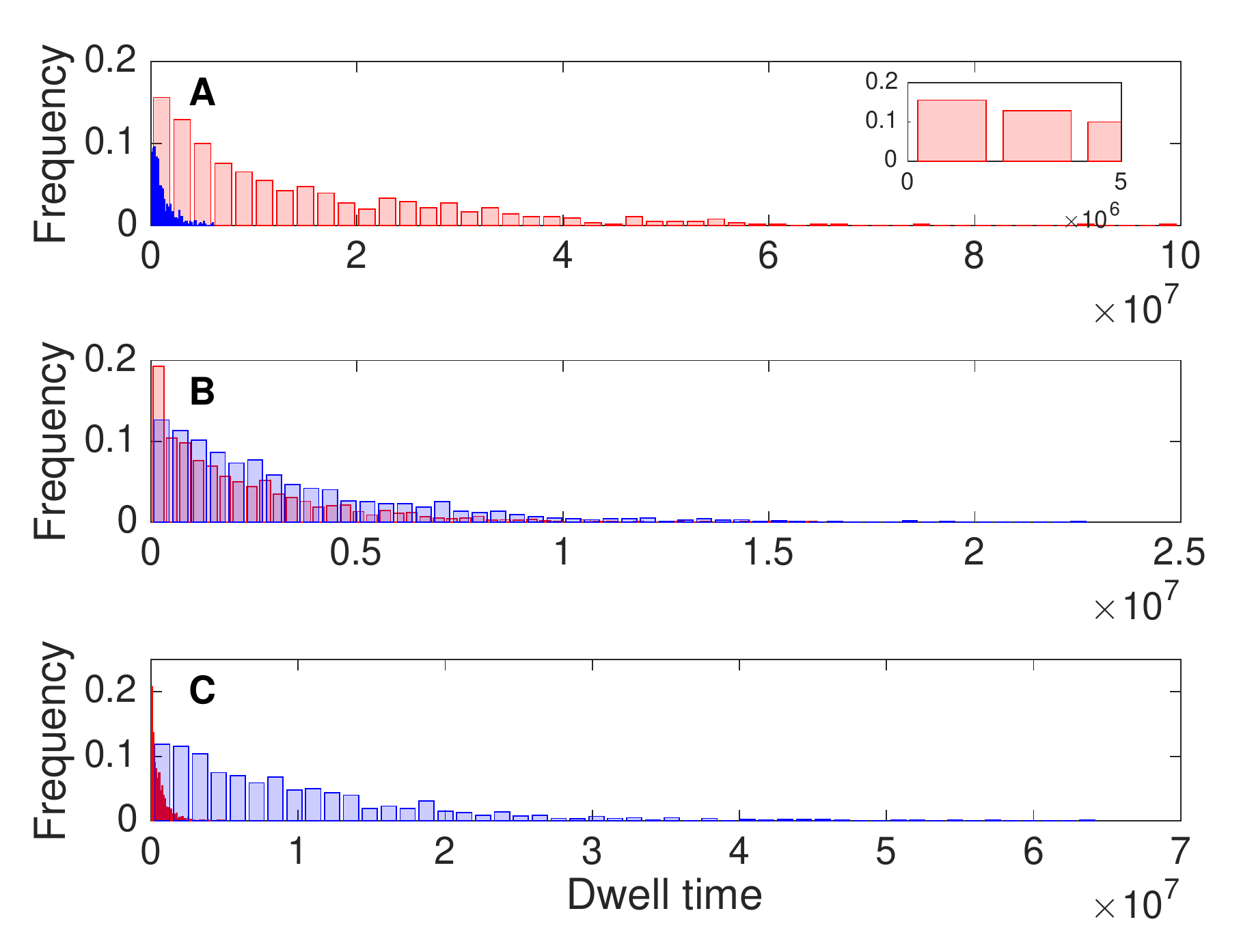}
\caption{ Histograms of the dwell times in each of the bistable states, calculated from the time series of the number of proteins in the bistable region, for $L=500$ sites, $\theta=21$, $d=0.002$, $k=0.8$, $\beta=0.015$ and $n=2$ and three different values of $\alpha$: (A) $\alpha=0.75$, (B) $\alpha=0.77$ and (C) $\alpha=0.79$. The red bars indicate the histogram of the dwell times in the upper stable state, and the blue bars the histogram of the dwell times in the lower stable state. The inset in panel (A) shows a zoom of the histogram for the upper stable state for low dwell times, to demonstrate that the smallest dwell times are larger than $0$.}\label{FigS_bistability_dwell_times}
\end{figure}

\begin{figure}[h]
\center
\includegraphics[width=\textwidth]{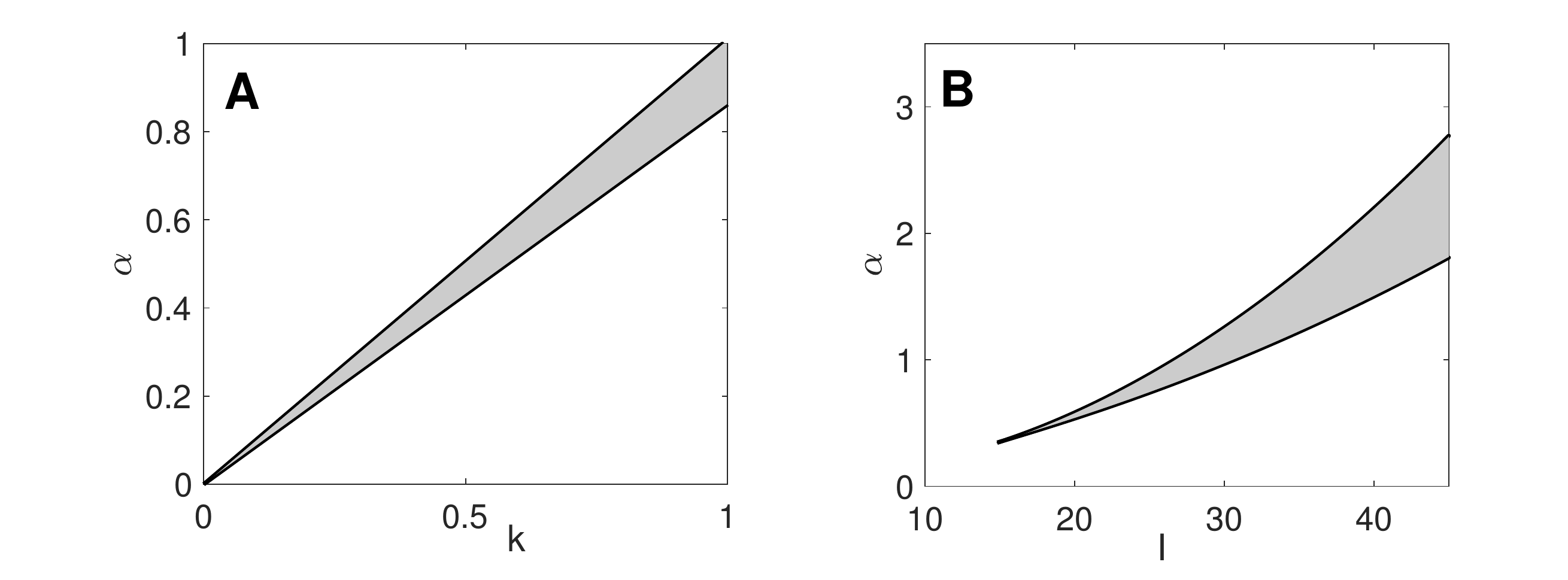}
\includegraphics[width=0.5\textwidth]{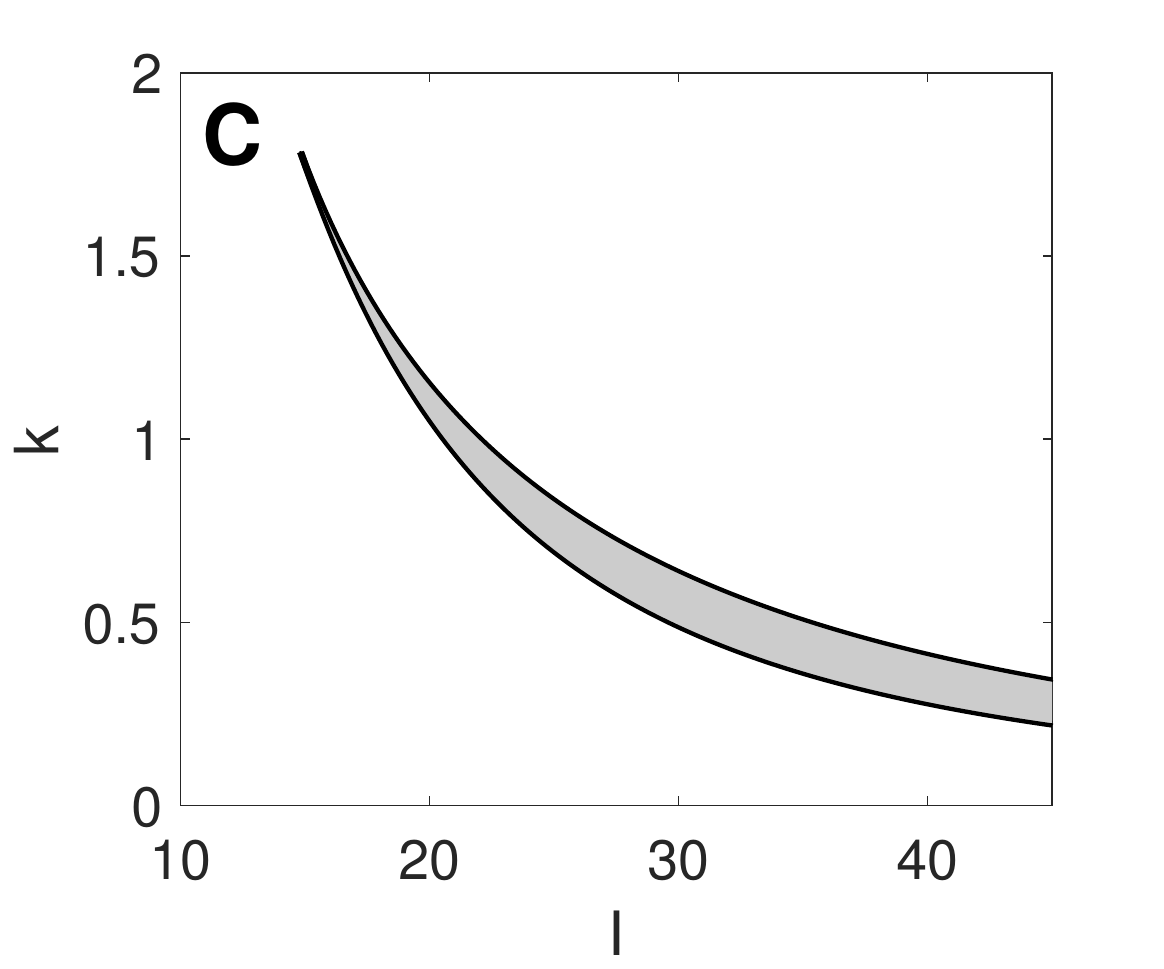}
\caption{Regions of bistability as predicted by the steady state theory shown as functions of  feedback ($I$), recycling ($k$) and  loading rates ($\alpha$). Bistable regions are shaded grey. 
 (A) $\alpha$-$I$ plane ($k=0.8$) (B) $\alpha$-$k$ plane ($I=24$) (C) $I$-$k$ plane  ($\alpha = 0.77$). Otherwise $L=500$, $n=2$,  $\beta = 0.015$.}\label{FigS-BS}
\end{figure}

\newpage